\documentstyle[proceedings]{kluwer}
\begin{opening}
\title{EVIDENCE FOR THE GALACTIC BAR FROM THE TWO COLOR PHOTOMETRY
OF THE BULGE RED CLUMP STARS}
\author{K. Z. Stanek}
\institute{Princeton University Observatory\\
Princeton, NJ 08544--1001}
\author{M. Mateo}
\institute{Department of Astronomy, University of Michigan \\
821 Dennison  Bldg., Ann Arbor, MI~48109--1090}
\author{A. Udalski}
\author{M. Szyma\'nski}
\author{J. Ka\L u\.zny}
\author{M. Kubiak}
\institute{Warsaw University Observatory \\
Al. Ujazdowskie 4, 00--478 Warszawa, Poland}
\author{W.~Krzemi\'nski}
\institute{Carnegie Observatories, Las Campanas Observatory \\
Casilla 601, La Serena, Chile}
\end{opening}

\runningtitle{GALACTIC BAR FROM PHOTOMETRY OF RED CLUMP STARS}

\begin{document}
\vspace{-0.45cm}
\section{INTRODUCTION}

The Optical Gravitational Lensing Experiment (OGLE, Udalski et al.~1994a;
Paczy\'nski et al.~1994b -- these proceedings; and references therein) is an
extensive photometric search for the rare cases of gravitational microlensing
of Galactic bulge stars by foreground objects.
It provides a huge data base (Szyma\'nski \& Udalski 1993), from which
color-magnitude diagrams have been compiled (Udalski et al.~1993, 1994b).
Here we discuss the use a of well-defined population of bulge red
clump stars to investigate the presence of the bar in our Galaxy.
The results of our earlier studies are described by Stanek et al.~(1994).

There is now a number of photometric and dynamical indications that the
Galaxy is barred (de Vaucouleurs 1964; Blitz \& Spergel 1991;
Binney et al.~1991; Whitelock \& Catchpole 1992; Weinberg 1992;
for a review see Blitz 1993).
The bar is clearly present in the COBE DIRBE data (Weiland et al.~1994),
which was used by Dwek et al.~(1994; also this proceedings) to constrain
a number of analytical bar models existing in the literature.

\begin{figure}[t]
\vspace{8cm}
\includegraphics{stanek_fig1.ps}
\caption{Positions in the Galactic coordinates of 19 fields discussed
in this report, for which the  $V-I$ color-magnitude diagrams
were obtained by the OGLE experiment (Udalski et al.~1993, 1994b).}
\end{figure}

\section{THE DATA}
Udalski et al.~(1993, 1994b) present color-magnitude diagrams (CMDs) of
19 fields in the direction of the Galactic bulge, which cover nearly 1.5
square degrees and contain about $8 \times 10^5$ stars. Fig.1 shows
the positions of all 19 fields in galactic coordinates.
As an example, the CMD  for one of the positive galactic longitude
fields (MM7-A) is shown in Fig.2, with main populations of stars
schematically illustrated. The part of the diagram dominated by the disk stars
was recently analyzed by Paczy\'nski et al.~(1994a). Here we use a
well-defined population of bulge red clump stars to investigate the presence
of the bar in our Galaxy. The results of our earlier studies
are presented by Stanek et al.~(1994).

To analyze the distribution of bulge red clump stars in a quantitative
manner, we define the extinction-insensitive $V_{_{V-I}}$ parameter
\begin{equation}
  V_{_{V-I}} \equiv V - 2.6 ~ (V-I),
\label{eq:free}
\end{equation}
where we use reddening law $ E_{_{V-I}} = A_{_V}/2.6 $.
The parameter $ V_{_{V-I}} $ has been defined so that if
$A_{_V}/E_{_{V-I}}$ is independent of location then for any particular star
its value is not affected by the unknown extinction (see Stanek
et al.~1994). The part of the CMD dominated by
bulge red clump stars is shown, for six out of all 19 observed
fields, in Fig.3. The fields were ordered from top-left to bottom-right by
decreasing galactic longitude $l$. Also shown are two lines corresponding
to the $V_{_{V-I}}$ values equal to 11.5 and 13.0.
It is clearly visible that the red clump stars from fields with larger $l$
group near  to the $V_{_{V-I}}=11.5$ line, while red clump stars from the
fields with smaller (negative) $l$ have, on average, larger values of this
parameter.

\begin{figure}[t]
\vspace{8cm}
\includegraphics{stanek_fig2.ps}
\caption{The $V-I$ color-magnitude diagram for stars in the
MM7-A field of the OGLE experiment (Udalski et al.~1993).
Schematically shown are main populations of stars: TOP -- Turn-Off Point,
RG -- Red Giants, RC -- Red Clump, SG -- Super-Giants and DMS -- Disk
Main-Sequence.}
\end{figure}
\begin{figure}[t]
\vspace{8cm}
\includegraphics{stanek_fig3.ps}
\caption{Region of the $V-I$ color-magnitude diagrams
dominated by bulge red clump stars for six out of all 19 observed fields,
ordered from top-left to bottom-right by decreasing galactic longitude $l$.
The two straight lines correspond to the value of extinction-insensitive
parameter $V_{_{V-I}}$ equal to 11.5 and 13.0 (Eq.1).}
\end{figure}
\begin{figure}[t]
\vspace{8cm}
\includegraphics{stanek_fig4.ps}
\caption{Plot of the $V_{_{V-I}}$ distributions for the same six fields
as in Fig.4. With the vertical dashed lines we mark the position
of the mode of the distributions.}
\end{figure}
\begin{figure}[t]
\vspace{8cm}
\includegraphics{stanek_fig5.ps}
\caption{Correlation between the mode of the $V_{_{V-I}}$ distribution
and the Galactic longitude $l$ for all observed fields.}
\end{figure}

To quantify the effect observed in Fig.3,
for all 19 fields we select the region of the CMD
clearly dominated by the bulge red clump stars:
\begin{equation}
1.4  < V-I ~~;~~ 10.5 < V_{_{V-I}} < 14.0
\label{eq:select}
\end{equation}
For every field all stars that satisfied the inequalities
(\ref{eq:select}) were counted in bins
of $ \Delta V_{_{V-I}} = 0.05 $. The result appears
in Fig.4, where we show the number of stars as a function of $V_{_{V-I}}$
for the same six fields as in Fig.3.
The distributions shown in Fig.4 are similar in shape,
with red clump stars forming a pronounced peak.
There is however a clear shift between the distributions, in the sense
that stars from fields with bigger value of $l$ have on average smaller
values of $V_{_{V-I}}$ parameter. To quantify this shift, for every field
we found the mode of the $V_{_{V-I}}$ distribution (see Stanek et al.~1994).
The resulting plot of the mode as a function of galactic longitude $l$ for
every field is shown in Fig.5. There is a clear anti-correlation between
the $V_{_{V-I}}$ value of the mode and the Galactic longitude $l$ for a
given field that corresponds to decrease of $V_{_{V-I}}$ value of
$\sim0.04\;mag/\deg$.

\section{DISCUSSION}

In previous section we have shown that the distributions of bulge red clump
stars, observed in various fields, as a function of extinction-adjusted
apparent magnitude  are similar in shape but are systematically shifted.
This is likely due to the difference in the distance to the bulge red
clump stars in various fields, an indication that the bulge is not axially
symmetric. If we assume that the peaks of the $V_{_{V-I}}$ distributions
correspond to the stars lying at the major axis of the bar, we can obtain
the angle of inclination of the bar to the line of sight,
$\theta\approx45\deg$. For a very thin bar such an inclination angle
corresponds directly to the true inclination angle, but if the bar is
thick then the true inclination angle is smaller (Stanek et al.~1994).
It is possible to use the observed luminosity function
of red clump stars to put constraints on various models of the Galactic bar.
This have the advantage over the studies using surface brightness
measurements (Dwek et al.~1994) that the red clump stars provide us
with information about the depth of the bar along the line of sight. Such
information was recently used by Stanek~(1994) in his study of the magnitude
offset between gravitationally lensed stars and observed stars. The preliminary
results from the modelling of the Galactic bar by fitting the observed red
clump luminosity function (Stanek et al., in preparation) indicates that
the inclination of the Galactic bar to the line of sight is about
$\sim 20\deg$, but it is too early to say how robust this result is.

The presence of the bar in the Galaxy seems to be firmly established
by various authors and methods, but there are
still considerable differences as to details of the bar structure
or angle of inclination to the line of sight. We have shown that the red
clump stars can be very useful for investigating the Galactic bar, being both
numerous and relatively bright. We are now extending  the work
presented here, by incorporating more information provided by the red clump
region of CMDs, to test a variety of Galactic bar models
(Stanek et al., in preparation).

We would like to thank B.~Paczy\'nski, the PI of the OGLE project, for
encouragement, many stimulating discussions and comments. We acknowledge
comments from J.~E.~Rhoads, who read an earlier version of this report.
This work was supported with the NSF grants AST 9216494
and AST 9216830 and Polish KBN grants No 2-1173-9101 and BST438A/93.

\end{document}